\documentclass[a4paper,11pt]{article}%
\usepackage{graphics}
\usepackage{graphicx}
\usepackage{amssymb}
\usepackage{amsmath}
\usepackage{amsfonts}
\usepackage{algorithm}
\usepackage{subfigure}
\usepackage{times}
\usepackage{color}
\usepackage{amsthm}
\usepackage{amsfonts}
\usepackage[left=2.5cm,top=2.5cm,right=2.5cm,bottom=2.5cm]{geometry}
\usepackage{epstopdf}
\usepackage{pdfpages}
\usepackage{fullpage}
\usepackage{nohyperref}
\usepackage{url}%
\providecommand{\U}[1]{\protect\rule{.1in}{.1in}}
\makeatletter
\def\tagform@#1{\maketag@@@{\ignorespaces#1\unskip\@@italiccorr}}

\numberwithin{equation}{section}
\graphicspath{{figures/}}
\makeatother

\providecommand{\U}[1]{\protect\rule{.1in}{.1in}}

\begin{document}

\author{Ariel Navon\thanks{{Faculty of Engineering, Bar Ilan University, Israel{}}{}.
navon2@gmail.com.}, Yosi Keller\thanks{{Faculty of Engineering, Bar Ilan
University, Israel{}}{}. yosi.keller@gmail.com.}}
\date{}
\title{Financial Time Series Prediction using Deep Learning}
\maketitle

\begin{abstract}
In this work we present a data-driven end-to-end Deep Learning approach for
time series prediction, applied to financial time series. A Deep Learning
scheme is derived to predict the temporal trends of stocks and ETFs in NYSE or
NASDAQ. Our approach is based on a neural network (NN) that is applied to raw
financial data inputs, and is trained to predict the temporal trends of stocks
and ETFs. In order to handle commission-based trading, we derive an investment
strategy that utilizes the probabilistic outputs of the NN, and optimizes the
average return. The proposed scheme is shown to provide statistically
significant accurate predictions of financial market trends, and the
investment strategy is shown to be profitable under this challenging setup.
The performance compares favorably with contemporary benchmarks along
two-years of back-testing.

\end{abstract}

\section{Introduction}

Time series analysis is of major importance in a gamut of research topics, and
many engineering issues. This relates to analyzing time series data for
estimating meaningful statistics and pattern characteristics of sequential
data. In particular, the forecasting of future values based on previously
observed measurements. For instance, given the samples of a discrete signal
$\mathbf{x}(n)$, $n\in%
%TCIMACRO{\U{2124} }%
%BeginExpansion
\mathbb{Z}
%EndExpansion
$, the forecasting task aims to estimate $x(n+T)$, where $T>0$. Alternatively,
we aim to estimate other parameters of $\mathbf{x}(n^{\prime})$, $n^{\prime
}>n,$ such as the standard deviation of $\mathbf{x}(n^{\prime})$ denoted as
$\sigma(\mathbf{x}(n^{\prime}))$, or future time series trends,
\begin{equation}
\hat{y}(n|\mathbf{x},T)=sign(\hat{x}(n+T)-x(n)).
\end{equation}
A loss function, also denoted as a cost function,%
\[
\mathbb{L}(n)=\mathbb{L}(y(n),\hat{y}(n))
\]
is defined to quantify the accuracy of the resulting prediction $\hat{y}%
(n),$ with respect to the ground-truth $y(n)$.

%\textbf{Ariel: Yosi - should not we define more formally y(n) before reffering to it ?}

%\textbf{Ariel: Another issue - The references seems to be out of order. }

Numerous works studied time series data, by applying statistical approaches.
The Kalman Filter \cite{kalman1960} and Auto-Regressive (AR) Models are
seminal statistical approaches, where the ARMA \cite{durbin1959efficient} and
ARIMA \cite{box1970distribution} (Auto-Regressive Integrated Moving Average)
models are further generalizations of the AR, derived by combining the AR
models with the moving-average (MA) models. Other common AR schemes is the
ARCH (Autoregressive Conditional Heteroskedasticity), GARCH
\cite{bollerslev1986generalized} (Generalized ARCH) and the ARIMA algorithms,
and are often applied to financial series forecasting.

Financial time series analysis deals with the extraction of underlying
features to analyze and predict the temporal dynamics of financial assets. Due
to the inherent uncertainty and non-analytic structure of financial markets
\cite{tsay2005analysis}, the task proved to be challenging, where classical
linear statistical methods such as the ARIMA model, and statistical machine
learning (ML) models have been widely applied
\cite{ahmed2010empirical,bontempi2013machine}.

%\textbf{Ariel: I had added here a refernce to general DL overview as you suggested:}

Deep-Learning (DL) \cite{lecun2015deep} approaches that relate to computational algorithms using
artificial neural networks with a large number of layers,\ allows to directly
analyze raw data measurements without having to encode the measurements in
task specific representations. In this work we propose to harness DL networks
to efficiently learn complex non-linear models directly from the raw data (an
"end-to-end" approach), for the prediction the financial market trends. We
focus on direct prediction of Standard \& Poor's 500 (denoted as: "S\&P500")
index and assets trends based on raw-form data of equities rates, and utilize
probability estimates given by the trained neural network.

In particular, we propose the following contributions:

\textbf{First}, we present an end-to-end learning scheme based on the raw
rates data of assets, in contrast to previous financial forecasting schemes
where the analysis utilized multiple engineered features
\cite{dixon2016classification}.

\textbf{Second}, we propose a trading strategy that utilizes the probabilistic
predictions of the neural network model, to determine the entry and exit
points of a trade.

\textbf{Last}, the proposed trading system is shown to be profitable while
outperforming baseline schemes, as demonstrated in a challenging realistic
commissions charged trading environment.

%\textbf{Ariel: I had deleted here the references to the deleted chapters of Deep-Learning and basic
%terms in financial trading.}

The rest of this work is organized as follows. Prior works in time-series
forecasting and applying machine learning to financial data are surveyed at
Chapter \ref{chap:PriorArt}. The proposed Deep-Learning model
end-to-end approach for financial trends forecasting is introduced in Chapter
\ref{chap:DeepLearningBasedPredictionOfStockPriceTrends}, and the
probabilistic trading strategy is proposed in Chapter
\ref{chap:ProbablityBasedTradingStrategy}. These schemes are experimentally
verified in Chapter \ref{chap:Results}, by applying them to several stocks
assets from the S\&P 500 index, using realistic transaction costs.

\section{Background}

\label{chap:PriorArt}

%\textbf{Ariel: I had deleted here the reference to the missing section of classical time series approaches}

In this chapter we review previous works in time series forecasting in
general, and financial forecasting in particular. Section
\ref{sec:MachineLearningMethodsForFinancialForecasting} surveys
machine-learning schemes for financial forecasting, while Deep-Learning schemes for
financial data are discussed in Section
\ref{sec:PreviousWorksonDeepLearningBasedTimeSeriesAnalysis}.

\subsection{Machine Learning Approaches For Financial Data Forecasting}

\label{sec:MachineLearningMethodsForFinancialForecasting}

Machine learning algorithms are often applied to time series forecasting in
general, and financial time series prediction. Kanas \cite{kanas2001comparing}
showed that the non-linearity of the models being used for time series
forecasting is of major importance. Thus, one of the most commonly applied
schemes is the K-Nearest neighbors (kNN). The kNN algorithm assumes a
similarity between time series sequences that occurred in the past, to future
sequences, and as such, the nearest-neighbors are used to yield the forecasts
of the kNN model. Ban et al. \cite{ban2013referential} applied kNN to
multi-asset data, utilizing a set of stocks sharing similar dynamics. Thus,
achieving less bias and improved resiliency to temporal fluctuations, than
those of a single stock.

Hidden Markov Models (HMM) are also commonly applied to financial time series
forecasting. A HMM encodes a finite-state machine with a fixed number of
non-observable states. These hidden variables are assumed to be related by a
Markov process allowing HMMs to be applied to temporal pattern recognition
tasks. Hassan \cite{hassan2005stock} applied HMMs to forecasting stock market,
by training an HMM model on a specific stock, and matching past temporal
patterns to current stocks patterns. The prediction is derived by
extrapolating current prices based on past events. Decision Trees
\cite{lai2009evolving} and SVM \cite{tay2001application,kim2003financial} were
also applied to time series forecasting.

\subsection{Financial Time Series Analysis Using Deep Learning}

\label{sec:PreviousWorksonDeepLearningBasedTimeSeriesAnalysis}

%\textbf{Ariel: should we delete the (DL) as it was stated already at the abstract?}

Deep Learning (DL) techniques were successfully applied to a gamut of problems
such as computer vision \cite{krizhevsky2012imagenet,he2016deep}, automatic
speech recognition \cite{hannun2014deep,graves2014towards}, natural language
processing
\cite{collobert2008unified,melamud2016context2vec,ben2016semisupervised},
handwriting recognition \cite{graves2012supervised}, and bio-informatics
\cite{min2016deep}, to name a few, outperforming contemporary state-of-the-art
schemes. Yet, the use of DL in financial time-series data is not widespread.
Nevertheless, some DL schemes were applied to financial data, utilizing
text-based classification, portfolio optimization, volatility prediction and
price-based classification.

Ronnqvist \cite{ronnqvist2016bank} proposed a DL\ approach for estimating
financial risk based on the news articles. Publicly available textual data was
used to quantify the level of banking-related reports, and a classifier was
trained to classify a given sentence as distress or tranquility. For that two
NNs\ were applied, the first aims to reduce dimensionality by learning a
semantic representation, while the other NN is trained to classify the learned
representation of each sentence. Fehrer and Feuerriegel
\cite{fehrer2015improving} trained a text-based classifier to predict German
stock returns based on news headlines, and reported 56\% accuracy on a
three-class prediction of the following trade day, without developing a
trading strategy. Ding \cite{ding2015deep} studied a similar topic by using
structured information extracted from headlines to predict daily S\&P 500 movements.

Portfolio optimization is the optimal dynamic selection of investment assets,
and was studied by Heaton \cite{heaton2016deep}, by trying to predict a
portfolio that will outperform the biotechnology index IBB. For that, an
autoencoder was trained, using the weekly return data of the IBB stocks
2012-2016. All stocks in the index were autoencoded, and the stocks found to
be most similar to the their autoencoded representation were chosen.

Xiong \cite{xiong2015deep} applied Long-Short-Term Memory (LSTM) neural
networks to model the S\&P500 volatility, using the Google stock domestic
trends as an indicators of the market volatility. Thus, reflecting
multi-parameters macro-economic status, as well as the public mood, and
outperforming benchmarks such as linear Ridge/Lasso and GARCH, with respect to
the mean average absolute error percentage. A deep neural network for
financial market trends prediction was proposed by Dixon
\cite{dixon2016classification} that trained a prediction model using multiple
financial instruments (including 43 commodity and forex assets), aiming to
classify the future trend as either positive, flat or negative. The dataset
consisted of aggregated feature training sets of all symbols, each encoded by
9895 engineered features and price differences. The neural net consists of
five fully connected layers and the model is shown to predict the instrument's
trend, while ignoring transaction costs.

\section{Deep Learning Prediction of Stock Price Trends}

\label{chap:DeepLearningBasedPredictionOfStockPriceTrends}

In this work we aim to derive a DL based prediction model to forecast the
future trends of financial assets, based on the raw data. The proposed scheme
is a dynamic probabilistic estimation of the asset price trend, that can be
applied to active trading, by entering either a long or short position, while
determining the exit point given an open position is introduced in Chapter
\ref{chap:ProbablityBasedTradingStrategy}.

The term \textit{long trade} refers to the operation of buying an instrument for the sake of
later selling it in a higher price. Thus, it is used when a positive trend is
expected, and its potential loss is limited to the cost of the trade. A \textit{short trade} is opened when a price
decline is expected, by first borrowing a financial instrument, and later
closed by buying back the instrument originally borrowed. The losses from
short positions are unbounded, as future prices are unbounded. The terms
\textit{buy} and \textit{long} are used often interchangeably, as well as the
terms \textit{sell} and \textit{short}.

Let $\mathbf{x}(n)$, $n\in%
%TCIMACRO{\U{2124} }%
%BeginExpansion
\mathbb{Z}
%EndExpansion
$ be a discrete financial time-series signal, such as the historic closing
prices of an asset, and $y(n|\mathbf{x},T)\in\left\{  -1,1\right\}  $ is the
temporal trend such that%
\begin{equation}
y(n)=\left\{
\begin{array}
[c]{@{}ll}%
1, & \text{if}\ \mathrm{x(n+T)>x(n)}\ \text{ positive trend}\\
-1, & \text{if}\ \mathrm{x(n+T)\leqslant{x(n)}}\ \text{ negative trend}%
\end{array}
\right.  \label{equ:trend}%
\end{equation}
where $T>0$ is the prediction interval.

Let $\widehat{x}(n)$ and $\hat{y}(n)$ be the estimated signal and trend,
respectively. We aim to minimize the softmax loss function $\mathbb{L}%
(n)=\mathbb{L}(y(n),\hat{y}(n))$ trained using the price logs data
$\mathbf{x}(n)$%
\begin{equation}
\mathbb{P}(y=j|\mathbf{x}(n))=\frac{e^{Z_{j}}}{\sum_{k=1}^{K}e^{Z_{k}}}%
\end{equation}
for $j\in{1,..,K}$, where $\mathbf{Z_{k}}$ are the inputs to the softmax layer
of the neural network. For the Up/Down two-classes classification problem,
$K=2$, and the logistic loss is given by%
\begin{equation}
\mathbb{L}=-log(\frac{e^{Z_{i}}}{\sum_{k=1}^{2}e^{Z_{k}}})
\end{equation}
where $i$ is the ground truth label $y(n)$. The proposed scheme utilizes both
the \textit{hard} and \textit{soft} estimates of $y(n)$, and we denote the
soft estimate $p(y|\mathbf{x}).$

\subsection{Deep Learning-based Prediction of Price Trends}

\label{sec:System}

\begin{figure}[tbh]
\centering
\includegraphics[width=0.99\linewidth]{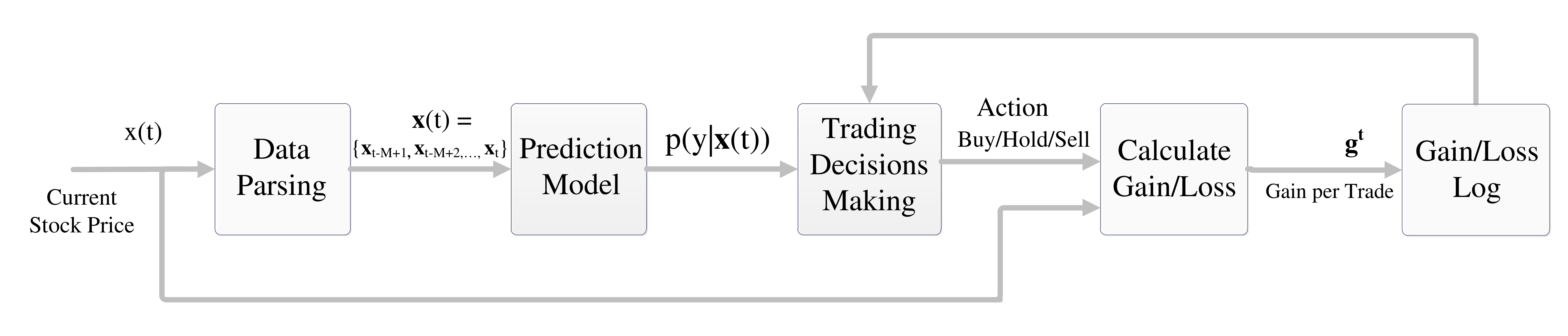}
%{DrawingsAriel/SystemDrawing}
\caption{Schematic Illustration of our Deep-Learning based financial trends
prediction system}%
\label{fig:System}%
\end{figure}

The proposed scheme, depicted in Fig. \ref{fig:System}, consists of two phases. The first, detailed in Section
\ref{sec:Model}, aims to predict the signal's trend $y(n)$ and corresponding
probability $p(y|\mathbf{x})$, while the second, discussed in Chapter
\ref{chap:ProbablityBasedTradingStrategy} applies the predicted trend to derive an investment
strategy, operating in a commissions-charged trading environment, where in each time step one can
either buy/hold/sell the asset. We start by preprocessing the price history
input data $\mathbf{x}(n)$ into a $M$-length features vector, consisting of
sequential normalized price values, and the temporal gain is given by
\begin{equation}
\mathbf{g}^{n}=\hat{y}(n)\cdot(x(n+T)-x(n)). \label{equ:temporal gain}%
\end{equation}

\subsection{Deep Learning Model for Price Trends Prediction}

\label{sec:Model}

\begin{figure}[tbh]
\centering
\includegraphics[width=0.99\linewidth]{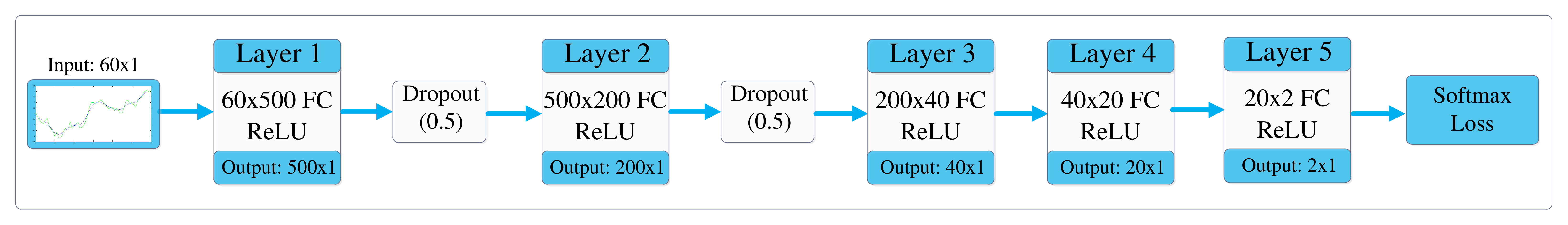}
%{DrawingsAriel/SystemDrawing}
\caption{The neural network applied to predict the price trends.}%
\label{fig:Net_Experimental_v1}%
\end{figure}

In order to predict the future stock price trend $\hat{y}(n)$ we apply a
classification neural network trained using the raw prices data of the
previous $M$ minutes%
\[
\hat{y}(n)=\hat{y}(n|x^{n-M+1},..,x^{n}).
\]
For that we applied the Neural Net depicted in Fig.
\ref{fig:Net_Experimental_v1}, and also experimented in utilizing
convolutional layers, though it did not yield significant accuracy
improvement, due to the low dimensional features space.

\subsection{Preprocessing}

Let $\mathbf{x}(n)$ be the input to the net consisting of closing price data
of S\&P500 assets in one-minute resolution. For each data point $\mathbf{x}%
(n)\in%
%TCIMACRO{\U{211d} }%
%BeginExpansion
\mathbb{R}
%EndExpansion
^{60}$, we used the raw closing prices of the previous $M=60$ minutes. 
The dataset was preprocessed by parsing the prices-data, where each sample
$S\left(  n\right)  $ encodes a particular minute during the trading hours of
a trading day. Each such sample is composed of its preceding 60 data points
$\mathbf{x}(n)$ and a label $y(n)$ representing the trend. Our approach
relates to intra-day trading, and was trained on data belonging to the same
trading day. In order to avoid irregular trading periods, we omitted trading
days that belong to earning publication periods, as well as omitting days with
partial trading hours. In order to avoid overfitting we used Dropout layers
\cite{hinton2012improving} after the first and second layers, as well as
early-stopping of the training, based on the validation set. The data set was
divided to temporally non-overlapping, training, validation and test sets.

\subsection{Labeling the Financial Dataset}

\label{sec:Labeling}

The duration of the prediction interval $T,$ relates to the correlation
between the features $\mathbf{x}(n)$ (past intra-day price values) and the
future values $y(n+T)$, that decreases as the prediction interval $T$ value
increases. In contrast, the financial price variations are more significant
for larger prediction intervals. This trade-off was resolved by studying
multiple datasets using $1\leq{T}\leq30$ minutes, where a different model was
trained for each dataset, and $T$ was chosen by cross-validation, as detailed
in\textbf{ }Table \ref{table:CrossValidation_SPY}.

\section{Probabilistic Trading Strategy}

\label{chap:ProbablityBasedTradingStrategy}

In order to apply the proposed DL\ prediction model to financial data, we
derive a buy-hold-sell probabilistic trading strategy, aiming to maximize the
overall cumulative return%
\begin{equation}
\mathbb{G}=\sum\limits_{n=1}^{n=N_{Tesn}}\mathbf{g}^{n}%
\end{equation}
along the back-testing period ($1\leq${$n$}$\leq{N_{Tesn}}$), where the
temporal gain $\mathbf{g}^{n}$ is given in Eq. \ref{equ:temporal gain}. The
main challenge in achieving trading profitability, is the\textit{ transactions
costs} $T_{C}$, charged per each transaction, as the average likelihood ratio%
\begin{equation}
R=\frac{E\{p(\hat{y}=y|\mathbf{x})\}}{E\{p(\hat{y}\neq{y|\mathbf{x}})\}}%
\end{equation}
between the probabilities of correct and incorrect predictions is typically
just a few percents. Moreover, the average intra-day volatility is relatively
small%
\begin{equation}
{T_{C}}\sim P(n+T_{intraday})-P(n)),
\end{equation}
where $T_{intraday}$ indicates a typical intra-day trading period. We
generalize the definition of $\mathbf{g}^{n}$ to $\mathbf{g}_{TC}^{n}$ taking
into account the transaction costs%
\begin{equation}
\mathbf{g}_{TC}^{n}=\hat{y}(n)\cdot(x(n+T)-x(n))-T_{C}, \label{eq:g_t}%
\end{equation}
such that we refer to a unified transaction cost for buy\&sell transactions
involved in a single trade.

We aim to identify the subset of the profitable transactions, denoted by
$\mathbf{\alpha}_{p(y|\mathbf{x})}\in\left\{  0,1\right\}  $, such that
$\mathbf{\alpha}_{p(y|\mathbf{x})}=1$ implies a profitable transaction, and
$\mathbf{\alpha}_{p(y|\mathbf{x})}=0,$ relates to a non-profitable one. The
cumulative gain is thus given by
\begin{equation}
\mathbb{G}=\sum\limits_{n=1}^{n=N_{Tesn}}(\mathbf{g}_{TC}^{n}\alpha
{_{p(y|\mathbf{x})}^{n}}).
\end{equation}
The proposed Probabilistic Trading Strategy utilizes the soft-information
$p(y|\mathbf{x})$ of the neural network output, used to estimate
$\mathbf{\alpha}_{p(y|\mathbf{x})}$. Section \ref{sec:SoftInfo} introduces the
trading strategy with respect to \textit{opening} a trade (when to
\textquotedblleft buy\textquotedblright), while Section \ref{sec:TradeLength}
discusses the closing of a trade (when to \textquotedblleft
sell\textquotedblright). Using long and short trades, the position could be
opened based on an either positive or negative expected price trends.

\subsection{Trade Opening Using Soft-Information}

\label{sec:SoftInfo}

The use of the soft-information $p(y|\mathbf{x})$ provided by the DL\ model
allows to select a subset of the trades with higher prediction accuracy,
improving the gain $\mathbb{G}$ while taking the commissions into account. For
that we consider the classification margin, such that%
\begin{equation}
\mathbf{\alpha}^{n}=\left\{
\begin{array}
[c]{@{}ll}%
1, & \text{if}\ \Vert p_{2}(y|x)-p_{1}(y|x)\Vert\geq T_{H}\text{\ }\\
0, & \text{if}\ \Vert p_{2}(y|\mathbf{x})-{p_{1}(y|\mathbf{x})}\Vert<{T_{H}}\
\end{array}
\right.  , \label{equ:thresh}%
\end{equation}
where ${T_{H}}$ is determined by cross-validation.

Due to the nonstationarity of the financial process, applying Eq.
\ref{equ:thresh} using a fixed threshold ${T_{H}}$, might prove inaccurate.
Hence, we propose to compute an adaptive threshold ${T_{H}}$ by cross
validation, where we apply ${T_{H}}$ to estimate the median gain over the $D$
previous data points. The value of $D$ was also estimated using
cross-validation and the validation set. We also considered a greedy approach
of choosing a threshold ${T_{H}}$ that maximizes the gain of the $D$ prior
points, but it was shown to be unstable and less accurate.

The use of the adaptive threshold ${T_{H}}$ allows to chose a subset of
profitable trading points. However, higher (and more realistic) trading
commissions (above 0.05\% of the transaction volume) require further\textbf{
}screening of the active trading points $\mathbf{\alpha}^{n}$. For that we
propose to avoid opening a new trades between $n\in\left[  n,n+N_{SafeLen}%
\right]  $ when the number of losses (negative gain trades) exceeds a
predefined threshold denoted as $N_{SafeTrigger}$%
\begin{equation}
\sum_{n-N_{SafeWin}+1}^{n}(\mathbf{g}_{TC}^{n}<0)\geq{N_{SafeTrigger}}%
\end{equation}

\begin{figure}[tbh]
\centering
\includegraphics[width=0.99\linewidth]{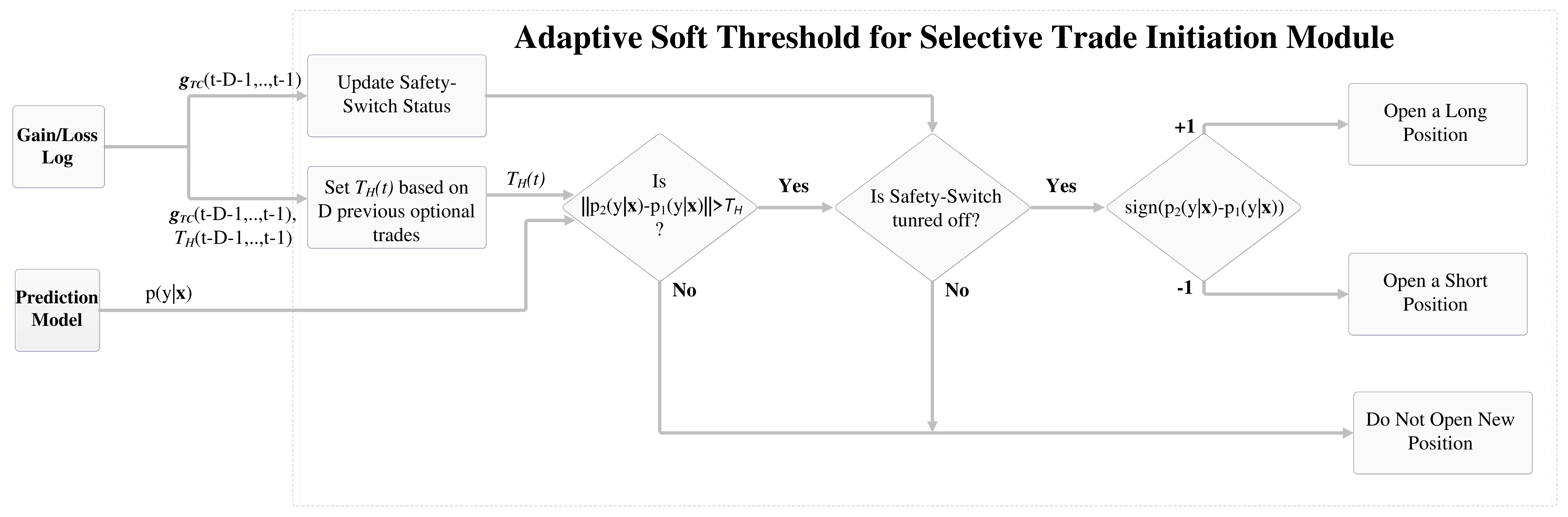}
\caption{The proposed soft-threshold based trade opening approach.}%
\label{fig:SoftTradingStrategy}%
\end{figure}

The last screening mechanism, targeted to experimentally avoid missclassified patterns, is noted as "safety-switch scheme".
The resulting trade opening scheme is depicted in Fig.
\ref{fig:SoftTradingStrategy}.

The safety-switch scheme significantly reduces the number of trade openings, down to $\sim$1\% of the overall number of the data points, as\textbf{ }depicted in
Fig. \ref{fig:170309_Dist_2}. Thus, choosing the more reliable predictions of
the net. We attribute that to the low certainty of the trend prediction DL
scheme that is able correctly classify only 53\% of the time slots, and to the high $T_C$ value comapring to the temporal gain value $g^n$ of a single trade.

\subsection{Setting the Trade-Length}

\label{sec:TradeLength}

We study the deal closure (\textquotedblleft sell\textquotedblright) event at
time $n_{c}>n_{o}$, where $n_{o}$ and $n_{c}$ are the deal opening and closure
times, respectively. Given the prediction interval $T$, mentioned in Section
\ref{sec:Labeling}, we considered three options for choosing the end point of
each trade. First, closing the trade after $T$ minutes regardless of the
predictions of $p(y|\mathbf{x})$, such that
\begin{equation}
n_{c}=n_{o}+T
\end{equation}
Second, closing the trade at $n_{c}$ such that the hard decision sequence
changes%
\begin{equation}
\hat{y}(n_{c})\neq{\hat{y}(n_{o}).} \label{equ: change}%
\end{equation}
Last, we tested waiting $T$ minutes after the hard decision sequence change as
in Eq. \ref{equ: change}. We found Eq. \ref{equ: change} to be the most
accurate, and it allows adaptive trading durations of varying lengths.

\section{Experimental results}

\label{chap:Results}

The applicability of the proposed DL scheme was studied by splitting the
dataset to temporally non-overlapping training, validation and test sets. The
test set consists of the data from June 23rd 2014 to June 22nd 2016. The
validation set is based on the data during the previous calendar year, from
June 23rd 2013 to June 22nd 2014, and all previous data was used as a training
set. Thus, our model does not account for changes in the market dynamics, and
implicitly assumes that a strategy learnt using the data up to 2013, can be
applied to trades in 2016.

We used the closing price data for trading, and chose assets with high trading
volumes, that are sufficiently liquid, such that the orders are always
executed on time (\textquotedblleft Spread\textquotedblright\ and
\textquotedblleft Slippage\textquotedblright\ effects are ignored). 

In each trade we invested an equal sum, and both the gain results and the transaction costs are measured in percentage of this sum, where we apply a combined commission rate for both
buy and sell actions. We report the cumulative gain over the two-years test period, that is the sum of
the gain of all active trades during that period of time.

\subsection{Experimental Setup}

\label{sec:ExperimentalSetup}

The proposed schemes were experimentally verified by using the market data of
the Standard \& Poor's 500 (\textquotedblleft S\&P 500\textquotedblright)
assets given in one-minute resolution, purchased from the QuantQoute market
data supplier \cite{QuantQuote}. For each asset analyzed, we utilized all
available history data of full-trading days, based on regular trading hours
(9:30AM-4:00PM), and ignoring off-hours trading data. We also omitted days
with partial trading hours, as well as earning-publication periods that are usually up to 2-3\% of the data points, and are given in the data set\textbf{.}

As we study intra-day trading, the data was divided to different trading days, where all features were derived from the current
trading day, with no overnight trades allowed. Each trading-day data is parsed into $\sim300$ different data
samples, and we used the raw closing price data in the preceding $M=60$
minutes, as features. Each input sequence was filtered by five taps long,
moving uniform averaging, and then normalized by reducing the mean, and
dividing by its standard deviation.

The training labels were set as in Eq. \ref{equ:trend}, and multiple models
were trained for varying $T$ values. For each asset we chose $T$ using
cross-validation, where common values range between $20-30$ minutes, as
depicted in Table \ref{table:CrossValidation_SPY}.

We applied over-sampling to balance the training and validation sets, such
that there will be an equal number of positive and negative training samples
for the trend prediction. The overall size of the data sets after parsing and
balancing depends upon the dates of available data for each asset, the model
parameters, and trading strategy. For the SPY ETF (detailed in Section
\ref{sec:ResultsSPY}) with $T=1$ there are $\sim1.2M$ samples before
balancing, divided to 967K training, 81K validation, and 164K test samples.
The test set was not balanced.

The prediction model is based on the neural network depicted in Fig.
\ref{fig:Net_Experimental_v1} that was trained using the MatConvNet
package \cite{vedaldi15matconvnet} with mini-batches of size 100, where the
learning rate was adaptively reduced, by a factor of 5, when observing a flat
error reduction of the validation set during training. The validation set was
also used for early-stoppage. The reliability soft-information provided by the
model is used by the proposed probabilistic trading strategy for detecting
high confidence trades.

\subsection{Implementation details}
The network architecture for our base NN model is shown in Fig.
\ref{fig:Net_Experimental_v1}. It has 5 fully-connected layers with 500, 200,
40, 20 and 2 ReLU activation units. Dropout follows the first and second
layers. The input is a 60 1 low-pass-filtered adjacent historical raw prices
values. We apply Stochastic-Gradient-Decent to train the models with 100 batch
size and 0.5 dropout rate. Initial learning rate is 0.001 and decreased by a
factor of 5 as validation errors stop decreasing, down to learning rate of
1e-7. All initial weights are sampled from Gaussian distribution with 0.01
standard deviation. We implemented the system in the MatConvNet framework. The
average inference time per input single price vector is 0.2ms on single Nvidia
GeForce GTX 780 GPU for the overall system framework.

\subsection{Active Trading of SPY ETF}

\label{sec:ResultsSPY}

We exemplify our schemes by applying the proposed active-trading to the SPY
ETF, that is the SPDR exchange-traded fund (ETF) - a Standard \& Poor's
Depositary Receipts, designed to track the S\&P 500 stock market index,
allowing investors to buy/sell the entire S\&P 500 index using a single
instrument. This is one of the ETFs with the largest trading volumes and
liquidity. The learning period $D$ and the position holding time $T$ were set
to $D=5$ and $T=28$, respectively using cross-validation, as shown at Table
\ref{table:CrossValidation_SPY}.

\begin{figure}[tbh]
\centering\includegraphics[width=0.99\linewidth]{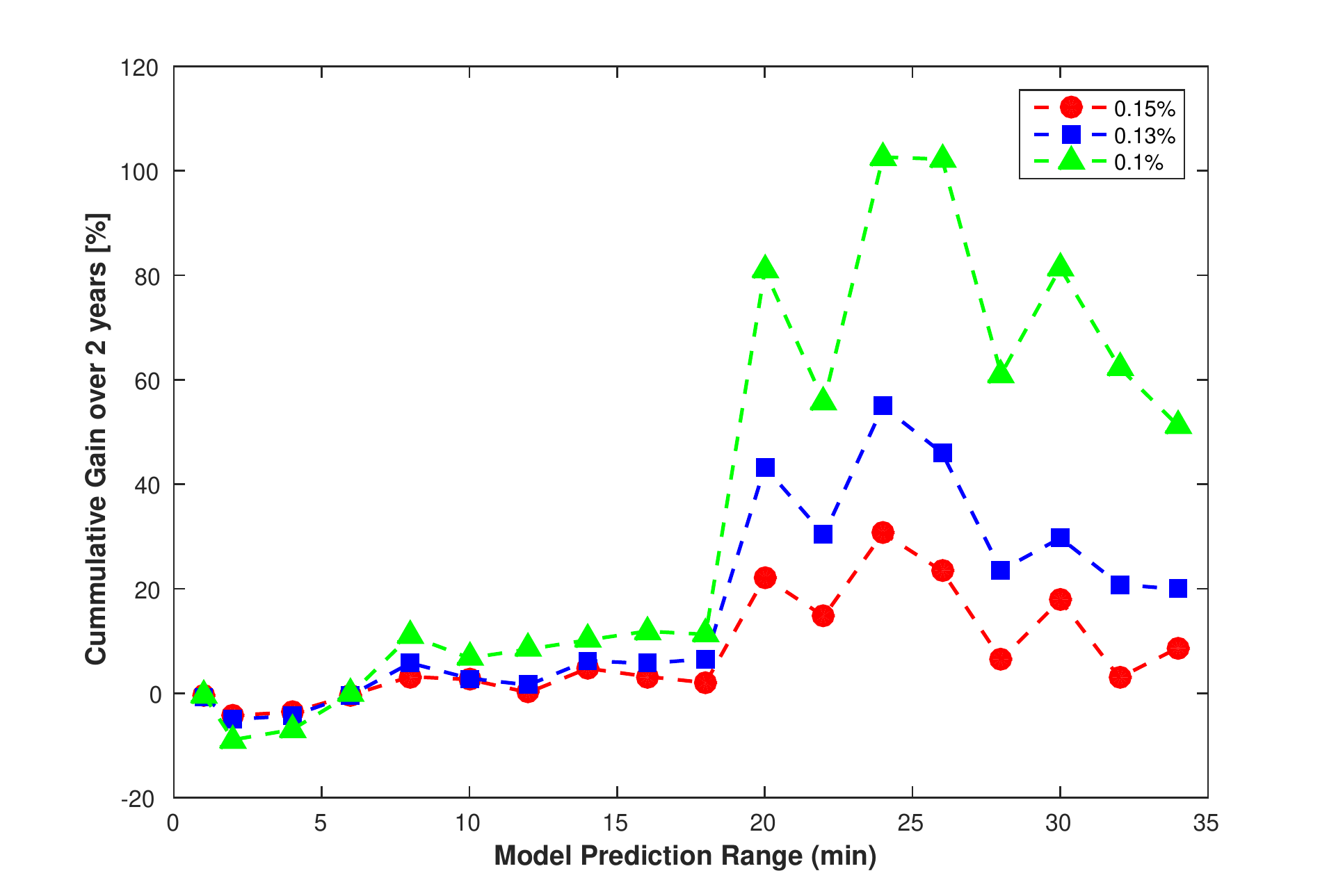}\caption{Cumulative
trading gain with SPY ETF over two years for different model prediction
ranges. We report the results for different commission rates.}%
\label{fig:160126_5minwin_GainsPerModel_differentComs}%
\end{figure}

During the testing period consisting of the last two years of our data
(23.6.2014-22.6.2016), the SPY price increased by 10.63\%. Figure
\ref{fig:160126_5minwin_GainsPerModel_differentComs} depicts the cumulative
trading gain over the test period for different models vs. the prediction
range $T$. We trained a different DL prediction models for each value of $T$,
and report the results for different transactions costs. It follows that the
DL model chosen by setting $D$ and $T$ according to the validation set (Table
\ref{table:CrossValidation_SPY}) yields a cumulative gain of 61.1\% over the
testing period for commision of 0.1\%.\begin{table}[h]
\begin{center}%
\begin{tabular}
[c]{|c|cccc|}\hline
& \textbf{D=1} & \textbf{D=5} & \textbf{D=10} & \textbf{D=50}\\\hline
\textbf{T=1} & 1.5 & \multicolumn{1}{|c}{-0.8} & \multicolumn{1}{|c}{-0.5} &
\multicolumn{1}{|c|}{0}\\\hline
\textbf{T=4} & 0.5 & \multicolumn{1}{|c}{0.1} & \multicolumn{1}{|c}{-0.8} &
\multicolumn{1}{|c|}{-3.3}\\\hline
\textbf{T=8} & 1.8 & \multicolumn{1}{|c}{0} & \multicolumn{1}{|c}{-0.6} &
\multicolumn{1}{|c|}{-2.5}\\\hline
\textbf{T=12} & 3.8 & \multicolumn{1}{|c}{2.7} & \multicolumn{1}{|c}{2.3} &
\multicolumn{1}{|c|}{-1.2}\\\hline
\textbf{T=16} & 0 & \multicolumn{1}{|c}{2.3} & \multicolumn{1}{|c}{2.1} &
\multicolumn{1}{|c|}{0.3}\\\hline
\textbf{T=20} & -3.7 & \multicolumn{1}{|c}{6.8} & \multicolumn{1}{|c}{3.1} &
\multicolumn{1}{|c|}{-7.8}\\\hline
\textbf{T=24} & 5.7 & \multicolumn{1}{|c}{20.2} & \multicolumn{1}{|c}{20.1} &
\multicolumn{1}{|c|}{-8.9}\\\hline
\textbf{T=28} & 2.9 & \multicolumn{1}{|c}{21.2} & \multicolumn{1}{|c}{18.9} &
\multicolumn{1}{|c|}{-3.7}\\\hline
\textbf{T=32} & 5.1 & \multicolumn{1}{|c}{16.9} & \multicolumn{1}{|c}{16.5} &
\multicolumn{1}{|c|}{-8.2}\\\hline
\end{tabular}
\end{center}
\caption{Cross validation results for the trade labeling length $T$ and
adaptive threshold window length $D$. We report the cumulative gain of the
proposed active-trading strategy over a one-year validation period, and a
commission of 0.08\%.}%
\label{table:CrossValidation_SPY}%
\end{table}

\begin{figure}[tbh]
\centering
\includegraphics[width=0.9\linewidth]{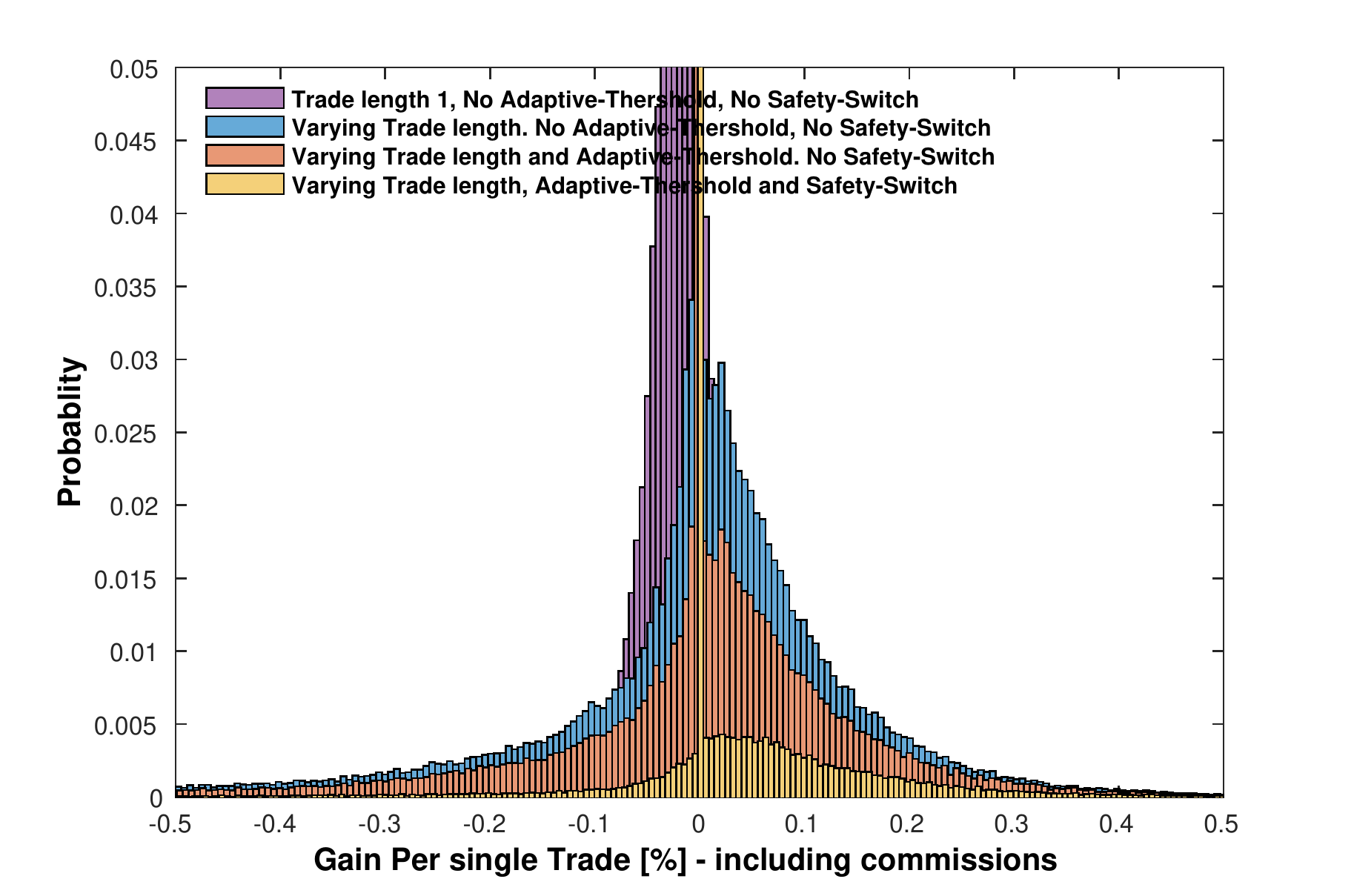}\caption{The
probability of returns-per-trade over the test period. We report the results
for different configurations of the proposed scheme.}%
\label{fig:170309_Dist_2}%
\end{figure}

We show the effect of the different phases of the proposed trading scheme in
Fig. \ref{fig:170309_Dist_2}, that depicts the histogram of the gain of
trades, when applying different components of the proposed scheme. For that we
compared the following schemes:\newline

\begin{itemize}
\item Fixed trading length of $T=1$. No adaptive-soft-threshold and
safety-switch mechanisms for selective trade initialization.

\item Varying trade length, by closing the trade when the model forecasting
sequence changes direction, without an adaptive-soft-threshold and a safety-switch.

\item Varying trade-length and adaptive-soft-thresholds, without a safety-switch.

\item Utilizing all of the proposes components as in Section
\ref{chap:ProbablityBasedTradingStrategy}.
\end{itemize}

It follows that the distribution of the proposed scheme outperforms the other
schemes, where a significant part of the distribution is concentrated at the
zero-gain bin, corresponding to the underlying assumption of the proposed
approach, as only trades probable to achieve a positive gain are chosen.

\subsection{Comparison to Benchmarks}

\label{sec:ComparisonToBenchmarks}

\begin{figure}[tbh]
\centering
\includegraphics[width=0.98\linewidth]{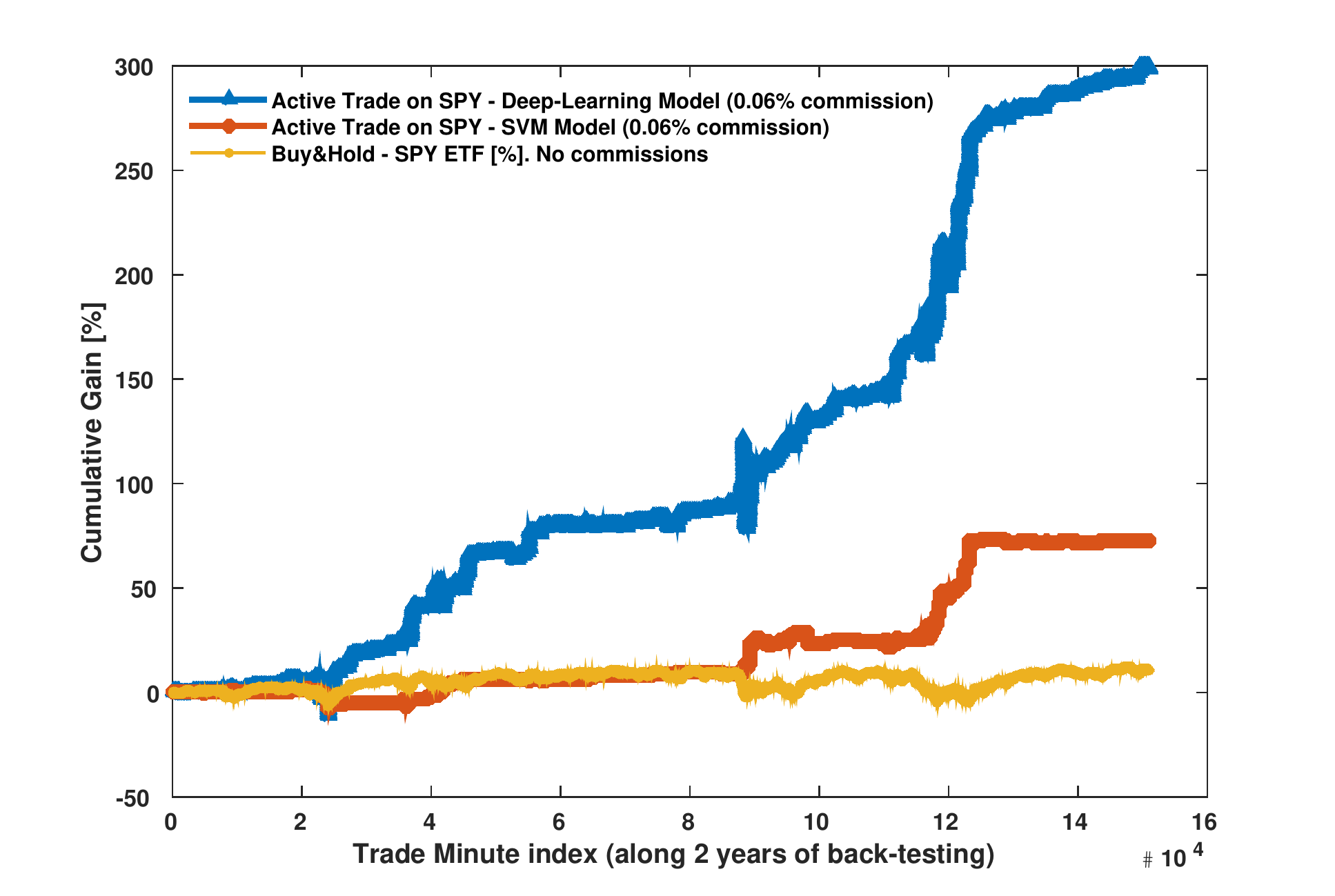}\caption{Cumulative
tradings gain. We report The performance of the proposed Deep-Learning based
scheme, a SVM prediction model utilizing the proposed probabilistic trading
strategy, and a buy-and-hold strategy.}%
\label{fig:160328_ExampleMinuteLevelGainTrade}%
\end{figure}

In order to evaluate the performance of our Deep-Learning based active trading
approach, we compared against the Support-Vectors-Machine (SVM) scheme, that
is considered a state-of-the-art classification scheme. The results over the
test period are presented in Fig. \ref{fig:160328_ExampleMinuteLevelGainTrade}%
, where in each time point we report the cumulative gain, starting with the
first test day. Both active-trading strategies (DL-based and SVM-based)
presented in Fig. \ref{fig:160328_ExampleMinuteLevelGainTrade}, include
transactions costs of 0.06\% (buy\&sell), where no commissions were applied
for the buy-and-hold strategy.

The Kernel SVM with a RBF kernels was implement by applying PCA to the parsed
data base, while preserving 95\% of the data variance, thus reducing the
dimensionality from 60 to 7. Higher values of the preserved variance lead to
inferior results. The classification probability estimate of the SVM was used
by the proposed Probabilistic Trading Strategy. These results were compared to
the baseline asset price change, corresponding to the buy-and-hold strategy.

\begin{figure}[tbh]
\centering\subfigure[]
{\includegraphics[width=0.5\linewidth]{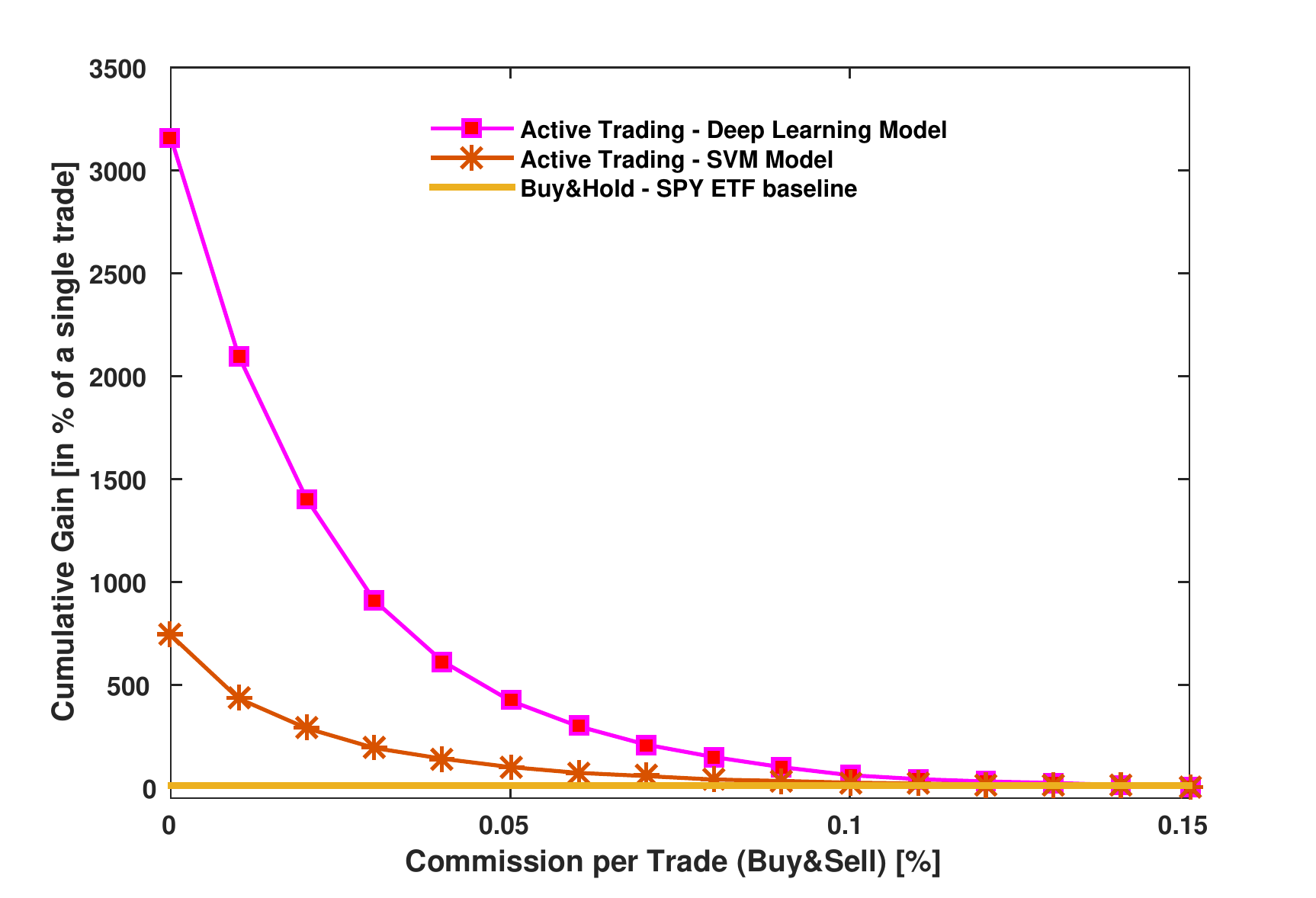}}%
\subfigure[]{\includegraphics[width=0.5\linewidth]{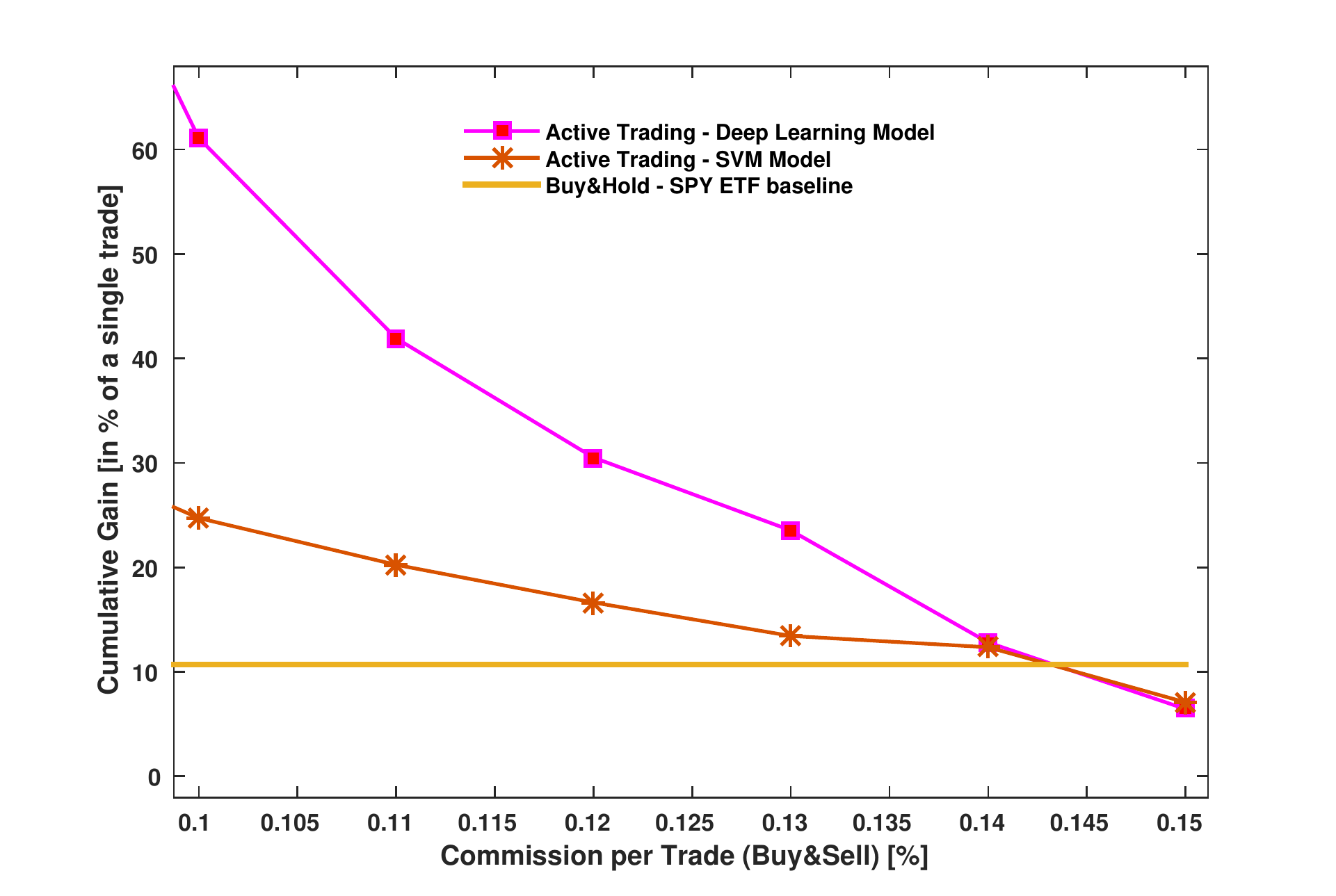}}\caption{Overall
cumulative gain over two years test period as a function of the transaction
cost. We report the results of the proposed probabilistic active trading
strategy for both the Deep-Learning model and the SVM model., as well as the
buy-and-hold investing strategy are also presented. (a) A overall range of
commissions. (b) The high commissions range.}%
\label{fig:170420_CumGain_DLvsSVMvsBuyHold}%
\end{figure}

Figure \ref{fig:170420_CumGain_DLvsSVMvsBuyHold} shows that the proposed DL
scheme outperforms the SVM-based model, and the asset-baseline benchmark.
However, as the transaction costs increase - the profitability of
commission-based models such as ours is reduced, leading to inferior
performance compared to the buy-and-hold passive strategy.

\subsection{Active Stocks Trading}

\label{sec:Results10}

In this section we apply the proposed scheme to nine stocks from the S\&P 500
index: INTC (Intel Corporation), AAPL (Apple Inc.), GOOGL (Google), BAC (Bank
of America Corporation), AMZN (Amazon.com Inc.), KO (The Coca-Cola Company), T
(AT\&T Inc.), JNJ (Johnson \& Johnson), BA (The Boeing Company). We applied
the same scheme as in the previous section, where a model was trained for each
stock separately. The test and validation periods are the same as those of the
SPY ETF.

\begin{figure}[tbh]
\centering
\includegraphics[width=0.9\linewidth]{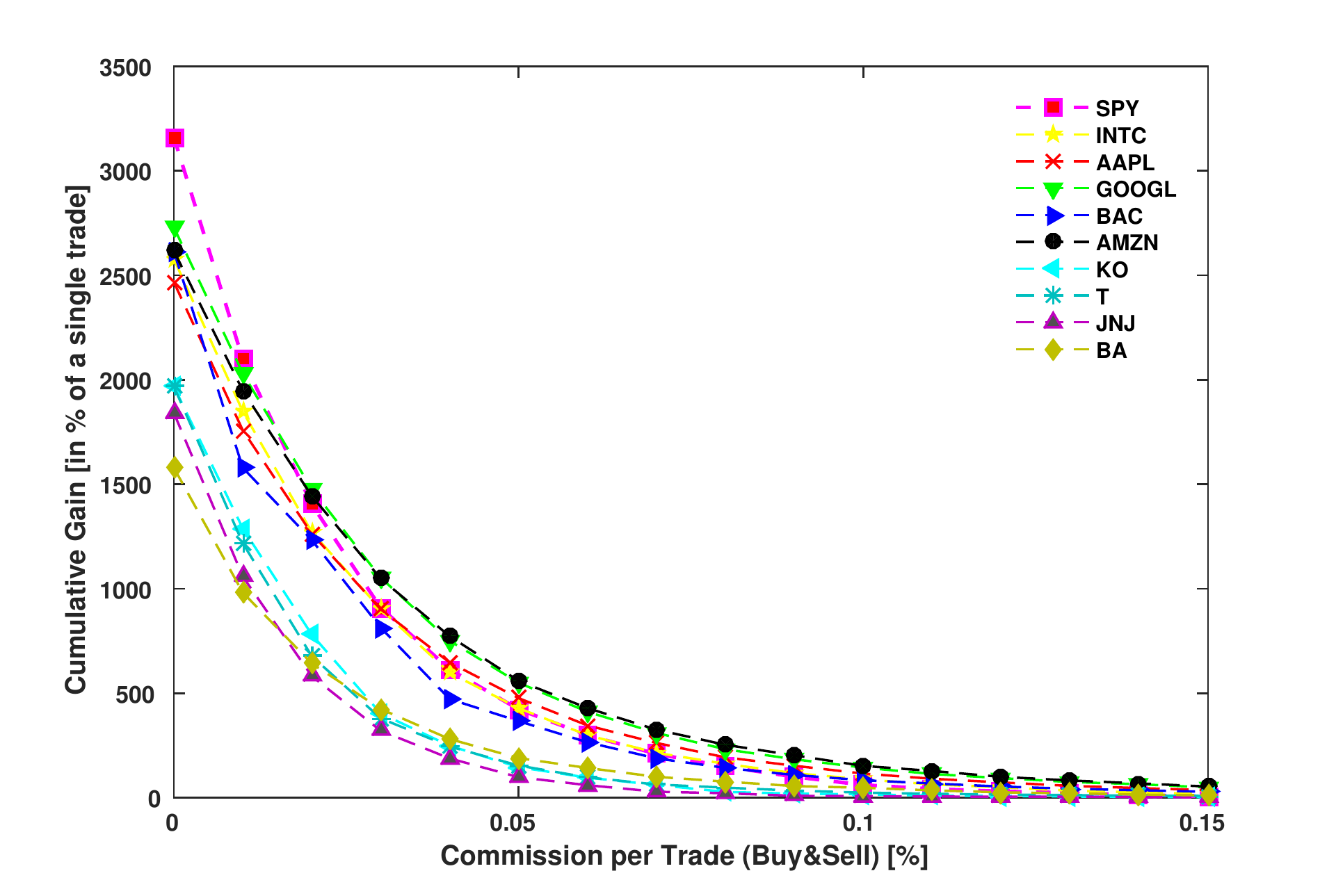}
\caption{Cumulative trading gain over two years of back-testing as a function
of the transaction cost (unified cost for buy\&sell). Each graph represents a
different instrument.}%
\label{fig:170409_CummulativeGainForDifferentInstruments_diffComs}%
\end{figure}

The cumulative gain over the two-years test period versus different
transaction commissions for the different instruments (including SPY) is shown
in Fig. \ref{fig:170409_CummulativeGainForDifferentInstruments_diffComs},
where it follows that the cumulative gains differs for different instruments,
and while positive results were achieved, they are strongly depended on the
transactions cost.

\begin{figure}[tbh]
\centering\subfigure[] {\includegraphics[width=0.5\linewidth]{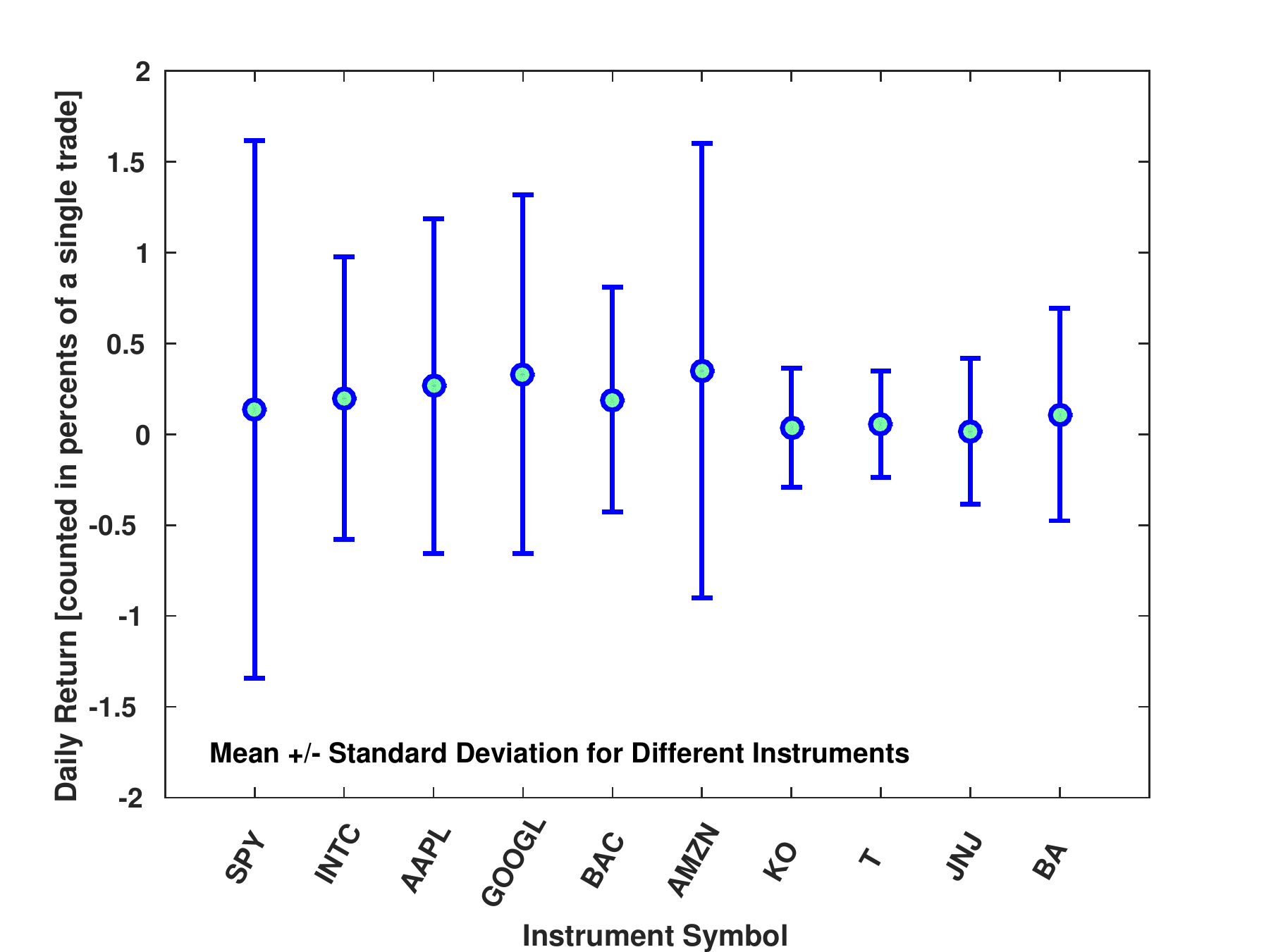}}\subfigure[]{\includegraphics[width=0.5\linewidth]{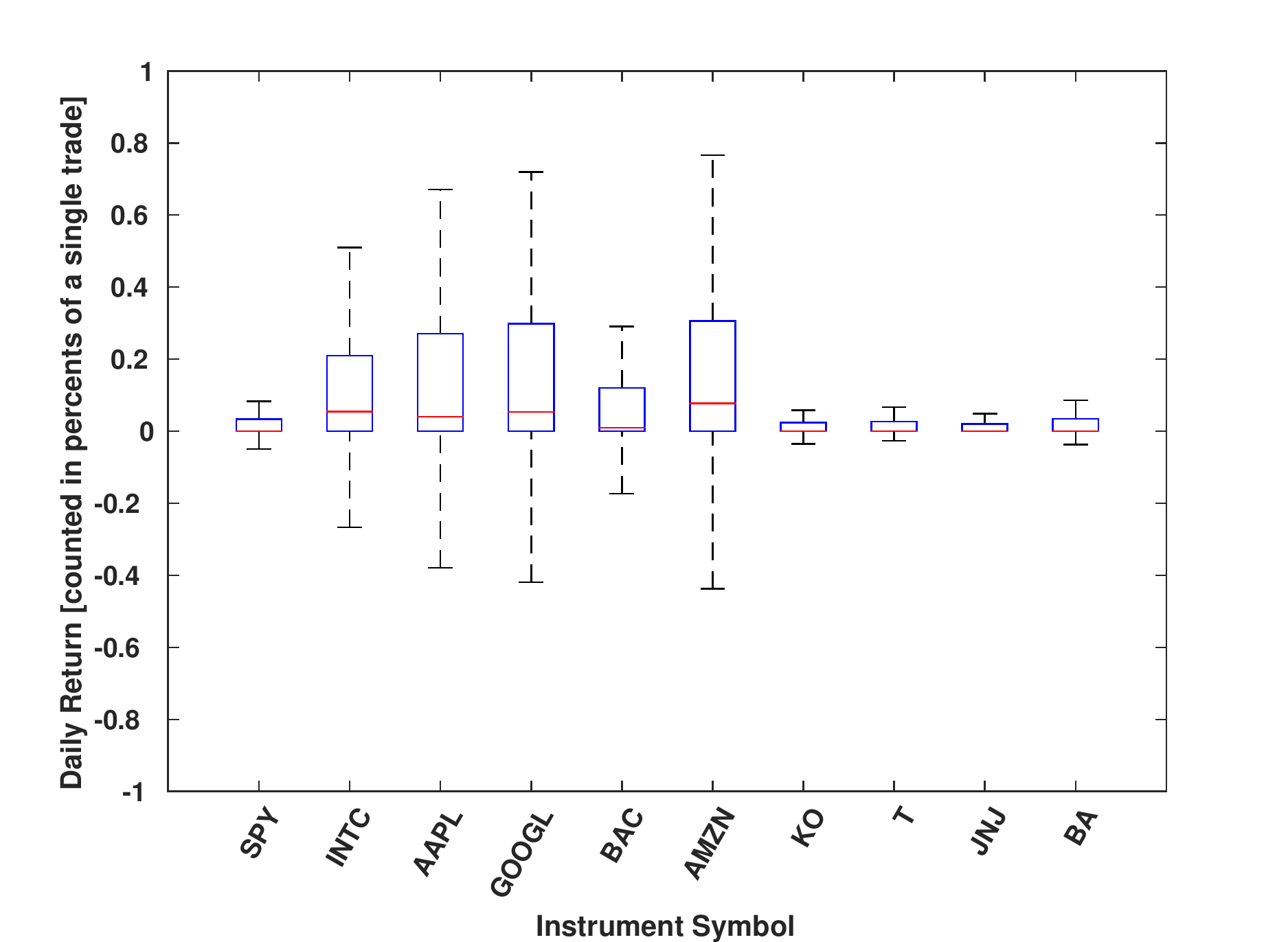}}\caption{Daily
return of the proposed Deep-Learning active trading strategy on different
instruments, and a commission of 0.1\%. (a) Mean standard deviation of the
daily returns of the different instruments. (b) the median, and 25\% and 75\%
percentiles.}%
\label{fig:Daily_Return}%
\end{figure}

Further analysis of is given in Fig. \ref{fig:Daily_Return}, where we study
the daily returns, assuming unified buy and sell commissions of 0.1\%. Figure
\ref{fig:Daily_Return}a reports the mean and standard deviation (volatility)
of the daily gain results of the different instruments, while Figure
\ref{fig:Daily_Return}b depicts the median, the 25\% and 75\% percentiles and
the region containing 99.3\% of the samples.

\begin{figure}[h!]
\centering
\includegraphics[width=0.85\linewidth]{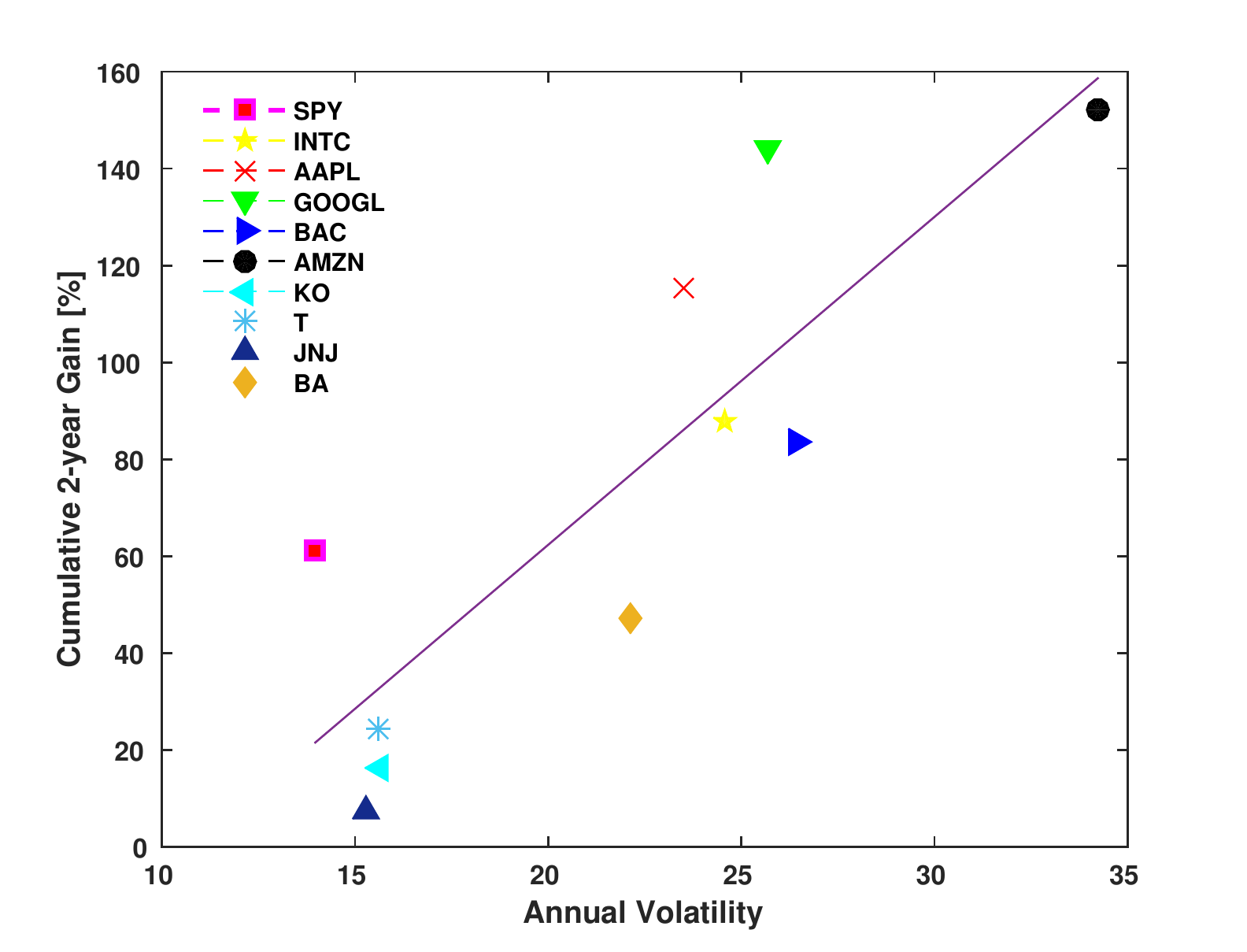}
\caption{Linear regression relation between the cumulative gain achieved by an
active-trading with different instruments as a function of the instruments
volatility.}%
\label{fig:170421_LinearRegression_VolatilyAndGain}%
\end{figure}

The volatility of a financial asset is an important measure of its risk \cite{InvestopediaTrading}. Table
\ref{table:CumGain_Baseline_Volatility} and Fig.
\ref{fig:170421_LinearRegression_VolatilyAndGain} report the cumulative
two-years returns of the DL approach for the ten instruments, and the
corresponding volatility. The volatility is computed as the annualized
historical volatility, given by%
\begin{equation}
\sigma_{Ann}=\sqrt{N_{minute}}\cdot{STD(\mathbf{g}_{TC})}%
\end{equation}
where $N_{minute}$ is the number of trading minutes per year and $g_{TC}$ is
the minute-level returns vector.

We also report the baseline instrument price. It follows that the
performance of the proposed active trading scheme is correlated with the
assets volatility, as exemplified in Fig.
\ref{fig:170421_LinearRegression_VolatilyAndGain}.

\begin{table}[h!]

\begin{center}%
\begin{tabular}
[c]{|c|ccc|}\hline
& \textbf{Active-Trade} & \textbf{Baseline} & \textbf{Ann. Volatility}\\\hline
\multicolumn{1}{|l|}{\textbf{SPY}} & 61.1 & \multicolumn{1}{|c}{10.6} &
\multicolumn{1}{|c|}{14.0}\\\hline
\multicolumn{1}{|l|}{\textbf{INTC}} & 87.7 & \multicolumn{1}{|c}{14.1} &
\multicolumn{1}{|c|}{24.6}\\\hline
\multicolumn{1}{|l|}{\textbf{AAPL}} & 115.4 & \multicolumn{1}{|c}{24.3} &
\multicolumn{1}{|c|}{23.5}\\\hline
\multicolumn{1}{|l|}{\textbf{GOOGL}} & 144.0 & \multicolumn{1}{|c}{35.1} &
\multicolumn{1}{|c|}{25.7}\\\hline
\multicolumn{1}{|l|}{\textbf{BAC}} & 83.6 & \multicolumn{1}{|c}{-11.1} &
\multicolumn{1}{|c|}{26.4}\\\hline
\multicolumn{1}{|l|}{\textbf{AMZN}} & 152.2 & \multicolumn{1}{|c}{86.1} &
\multicolumn{1}{|c|}{34.2}\\\hline
\multicolumn{1}{|l|}{\textbf{KO}} & 16.3 & \multicolumn{1}{|c}{18.5} &
\multicolumn{1}{|c|}{15.7}\\\hline
\multicolumn{1}{|l|}{\textbf{T}} & 24.2 & \multicolumn{1}{|c}{22.4} &
\multicolumn{1}{|c|}{15.6}\\\hline
\multicolumn{1}{|l|}{\textbf{JNJ}} & 7.4 & \multicolumn{1}{|c}{9.0} &
\multicolumn{1}{|c|}{15.3}\\\hline
\multicolumn{1}{|l|}{\textbf{BA}} & 47.4 & \multicolumn{1}{|c}{4.4} &
\multicolumn{1}{|c|}{22.2}\\\hline
\end{tabular}
\end{center}
\caption{Cumulative gain of the proposed active-trading strategy during
two-year test period (0.1\% commission), versus baseline instrument movement
and the annualized historical volatility.}%
\label{table:CumGain_Baseline_Volatility}%
\end{table}

\begin{figure}[h!]
\centering
\includegraphics[width=0.9\linewidth]{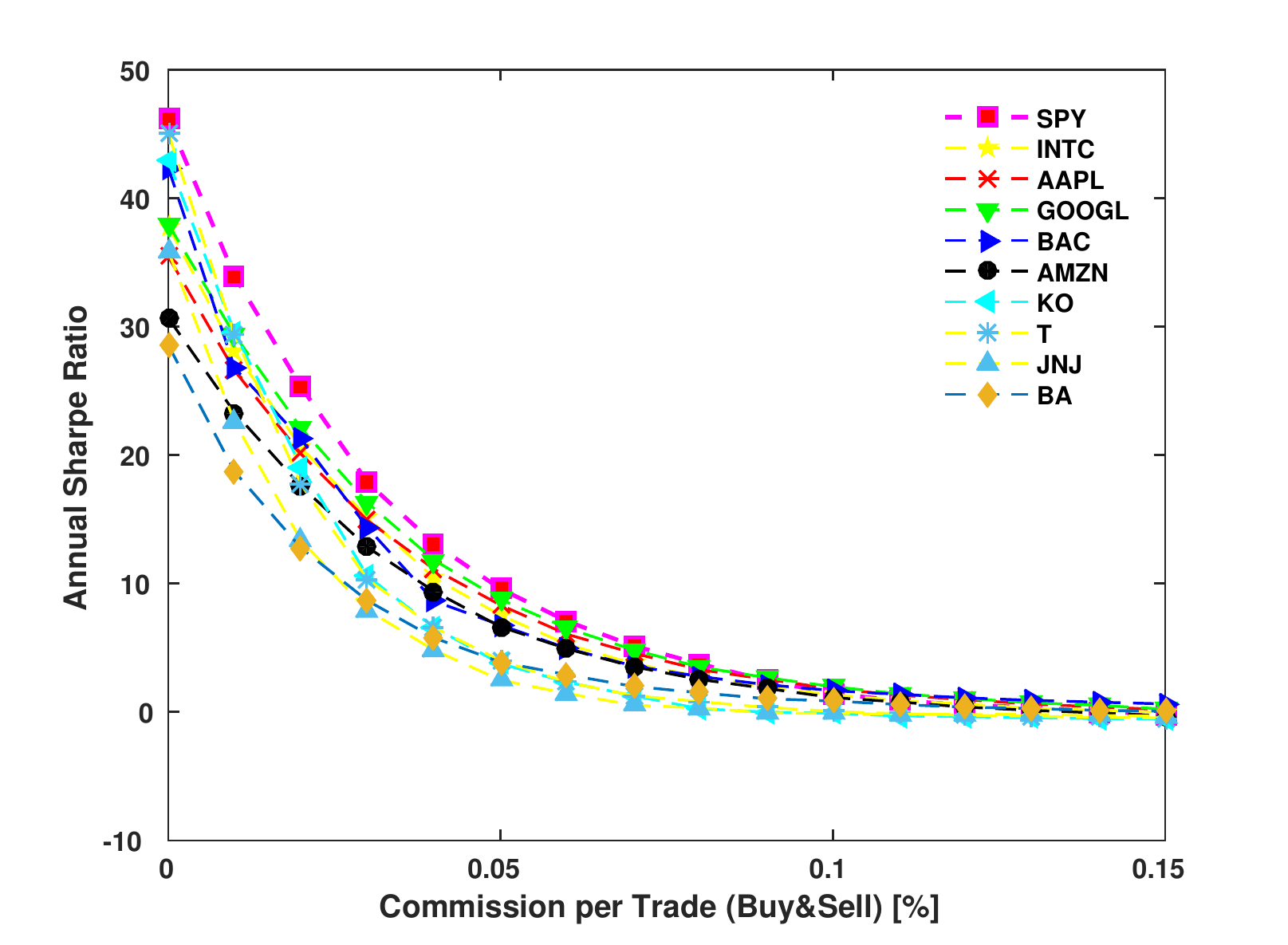}
\caption{Annual Sharpe ratio as function of trade commissions for different
instruments.}%
\label{fig:170404_AnnualSharpe_Ratio_as_function_of_Trade_Commissions}%
\end{figure}

Figure \ref{fig:170404_AnnualSharpe_Ratio_as_function_of_Trade_Commissions}
shows the annual Sharpe Ratio \cite{lo2002statistics} versus the commissions
for different assets using the proposed approach. As the Sharpe Ratio
\cite{sharpe1994sharpe} quantifies the risk-adjusted return, it follows that
it is strongly correlated with the commissions for all assets. The long term
Sharpe ratio of the S\&P 500 index that is commonly taken as 0.406, can be
used as a reference. Thus, for commissions of 0.07\% and lower, all assets
achieve Sharpe ratios above 0.5, and for commission of 0.1\%, three assets (T,
JNJ and KO) underperform. It should be noted that these three instruments have
the lowest volatility among all tested assets.

\section{Conclusions and future work}

\label{chap:conclusions}

In this work we presented a Deep Learning approach for financial time series
forecasting. For that, we propose an end-to-end Deep-Learning forecasting
model based on raw financial data inputs, contrary to common statistical
approaches to financial time series analysis which are based on engineered
features. Our approach is shown to produce statistically significant
forecasts, and we also derives a probabilistic active trading scheme achieving
profitability in a realistic commission charged trading environment. This
trading strategy utilizes the soft outputs of the DL to indicate the prediction
validity, and is shown to outperform both active-trading based on other
machine-learning algorithms, and the baseline buy-and-hold investing strategy.

Future work should include unifying the predicting model and the trading
strategy. The model could be dynamically updated based on the whole historical
data available at each point of time. These procedures are expected to produce
further improvement in the forecasting accuracy. Additional improvement could
be achieved by additional information from other sources, such as streaming
media reports. In particular, following Markowitz's portfolio theory,
optimizing a portfolio of assets (ETFs, stocks, etc.) should yield improved
results, in terms of lower volatility, and higher profitability.

\bibliographystyle{unsrt}

\bibliography{citations}

\end{document}